\documentclass{pnastwo}

\usepackage[dvips]{graphicx}

\usepackage{amssymb,amsfonts,amsmath,graphics, mathrsfs, color}

\oddsidemargin 0.2in \evensidemargin 0.2in \usepackage{graphicx}

\def\beq{\begin{equation}}
\def\beqa{\begin{eqnarray}}
\def\eeq{\end{equation}}
\def\eeqa{\end{eqnarray}}
\def\eqwithrates#1#2{\mathrel{\mathop{\rightleftharpoons}\limits^{#1}_{#2}}}
\begin{document}





\footcomment{SIB and CJ  performed the research.\\
SIB,  FH and CJ analyzed the data and wrote the paper.}
\conflictofinterest{The authors declare that they have no conflicts of interests.}








\title {Transcriptional pulsing and consequent stochasticity in gene expression}

























\author{Srividya Iyer-Biswas  \affil{1}{Ohio State University Department of Physics,
Woodruff Ave, Columbus, OH 43210}\thanks{To whom correspondence
should be addressed. E-mail:srividya@mps.ohio-state.edu}, F. Hayot \affil{2}{Mount Sinai
School of Medicine Department of Neurology, Levy Place, New York, NY
10029 }\and C. Jayaprakash \affil{1}{}}




\maketitle


\begin{article}

\begin{abstract}

Transcriptional pulsing has been observed in both prokaryotes and
eukaryotes and plays a crucial role in cell to cell variability of
protein and mRNA numbers. The issue is  how the time constants
associated with episodes of transcriptional bursting impact cellular
mRNA and protein distributions and reciprocally, to what extent
experimentally observed distributions can be attributed to
transcriptional pulsing. We address these questions by investigating
the exact time-dependent solution of the Master equation for a
transcriptional pulsing model of mRNA distributions. We find a
plethora of results: we show that, among others, bimodal and
long-tailed (power law) distributions occur in the steady state as
the rate constants are varied over biologically significant time
scales. Since steady state distributions may not be reached
experimentally we present results for the time evolution of the
distributions. Because cellular behavior is essentially determined
by proteins, we investigate  the effect of the different mRNA
distributions on the corresponding protein distributions. We
delineate the regimes of rate constants for which the protein
distribution mimics the mRNA distribution and those for which the
protein distribution deviates significantly from the mRNA distribution.

\end{abstract}











\section{Introduction}
Cell to cell variability in mRNA and protein numbers
is now recognized as a major aspect of cellular response to stimuli,
a variability which is hidden in cell population studies. The most
egregious example of the latter is provided in cases where a graded
average response hides the all-or-nothing behavior of single cells
\cite{fiering, hume, ko}. Variability of cellular response can have
many origins, which are generally classified as extrinsic and
intrinsic noise or fluctuations\cite{elowitz}. The source of intrinsic fluctuations is the
random occurrence of reactions that can lead to
variability for genetically identical cells in identical, fixed
environments. Extrinsic fluctuations can have multiple origins, such
as variations  from cell to
cell in the number of regulatory molecules, or signaling cascade components, or fluctuations
in cytoplasmic
and nuclear volumes. Many studies, both experimental and theoretical
 from bacteria to eukaryotes have
been undertaken to  disentangle intrinsic and extrinsic
 fluctuations
~\cite{elowitz,raser,blake, thattai, ozbudak, paulsson, raj, hu}.

Intrinsic fluctuations arise from either noisy
transcription or translation or both, the effects of which can be
measured in single cell mRNA and protein experiments. The simplest
model of protein number distributions is to consider  both
transcription
 and translation as Poisson
 processes~\cite{thattai}.
 Recent experimental studies of mRNA
 distributions have shown
strong evidence for transcriptional noise beyond what can be
described by a simple Poisson process. In particular,
transcriptional pulsing, where bursts of transcription alternate
with quiescent periods, has been observed in both prokaryotes and
eukaryotes. Raser and O'Shea \cite{raser}, who studied intrinsic and
extrinsic noise in \emph{Saccharomyces cerevisiae},  showed that
the noise associated with a particular promoter could be explained
in a transcriptional pulsing model and confirmed it by mutational
analysis. Transcriptional  bursts were recorded in
 \emph{E. coli}
\cite{golding} by following mRNA production in time,  and their
statistics computed.  Evidence  for a pulsing model of
transcription, obtained from fluorescent microscopy, has also been
presented for the expression of the discoidin Ia gene of
\emph{Dictyostelium}~ \cite{chubb}. Transcriptional bursts have as well been
detected in Chinese hamster ovary cells \cite{raj}.  In these
experiments the production of mRNA occurs in a sequence of bursts of
transcriptional activity separated by quiescent periods.
Transcriptional bursting, an intrinsically random phenomenon, thus
becomes an important element to consider when evaluating cell to
cell variability. One can predict that in many cases it will be a
significant part of overall noise, and most certainly of intrinsic
noise. It has been speculated that pulsatile mRNA production might
permit ``greater flexibility in transcriptional
decisions''~\cite{chubb}.  Cook et al.~\cite{cook} have argued that
different aspects of haploinsufficiency can be connected to time
scales associated with transcriptional bursting.

 Our study focusses on the consequences of transcriptional bursting in a
simplified model of transcription that has been the subject of many
studies and is believed to encapsulate the key features of
bursting~\cite{hume, raser, raj, golding, bose,kepler}. The complex
phenomena that can occur in transcription (chromatin remodeling,
enhanceosome formation, preinitiation assembly, etc.) are modeled
through positing two states of gene activity: an inactive state
where no transcription occurs, and an active one, in which
transcription occurs according to a Poisson process. The production
of mRNA is thus pulsatile: temporally there are periods of
inactivity interspersed with periods or bursts of transcriptional
activity. Qualitative features of this model were presented in
reference~\cite{hume}, and aspects of it relating to bursts
explored and discussed in reference \cite{golding}.
 Raj et al.~\cite{raj} provided a
steady state solution to the Master Equation of the transcriptional
model considered here, and analyzed it for some ranges of the rate
constants.   Given the range of time scales that can occur in
transcriptional processes in different organisms, it is imperative to
highlight the most significant behaviors that can arise in this model and investigate how these depend
on the many time scales. In this paper, we provide a comprehensive
analysis of a transcriptional pulsing model with an exact solution
to the {\it time-dependent} Master Equation for mRNA production. The
advantage of this model is that it is amenable to such an analytic
determination of the probability distribution of mRNA copy number as
a function of time. We find that the system exhibits a surprising
variety of distributions of mRNA number: this includes a bimodal
distribution with power-law behavior between the peaks that evolves
into a scale-invariant power-law distribution as we vary the rates
of activation and inactivation. In some systems the mRNA
distribution may not reach steady state, and it is therefore
necessary to determine the time evolution of the distributions and
characterize the time scales over which steady state is attained.
Our time-dependent analytic solution allows us to address these
issues in detail, revealing in particular that the mRNA lifetime
plays a key role in shaping the mRNA distribution. Cellular behavior is however determined by proteins and
not the corresponding mRNA. Therefore, an important question is to
what extent the protein distributions follow the mRNA distributions
obtained as a result of transcriptional pulsing. To answer this
question, we have performed numerical simulations of a model using
the Gillespie algorithm~\cite{gillespie} in which proteins are
produced in a birth-death process from mRNA.  When the protein
decay rates are much larger than the mRNA decay rate the protein
distributions reflect the mRNA distributions; when the protein
decays more slowly the protein distribution can be very different
from that of the parent mRNA, even in the steady state.
We delve deeply into the structure of the transcriptional bursting model, highlighting how the
shapes of mRNA
distributions depend on the ratios of time scales, determining over which time scales
distributions evolve to steady state, and stressing the importance of considering the full
distribution rather than characterizing it solely by its average and variance.
We thus provide an overview of possible behaviors which yield a framework for interpreting
experimental results on transcriptional bursting across prokaryotes
and eukaryotes.

\section{Results}
We study a model of transcriptional pulsing described by the
following reactions where $D$ and $D^*$ denote the gene in the
inactive and active states respectively: \beqa
D &\eqwithrates{c_f}{c_b}& D^* \\
D^*&\stackrel{k_b}{\longrightarrow}&D^*\,+\,M \\
M &\stackrel{k_d}{\longrightarrow}&\emptyset \,.
\eeqa
The first equation describes the switching ``on'' and ``off'' of the gene at
the rates $c_f$ and $c_b$ respectively. The second and third
equations describe transcription of the mRNA at a constant rate
$k_b$ in the active state and the subsequent degradation of the mRNA
at a rate $k_d$. We present results for $P(m,t)$, the
probability for the cell to contain $m$ mRNA molecules at a time $t$ that describes
cell-to-cell variability of mRNA copy number as a function of time.

\subsection{Time scales}~~
The steady state and temporal behaviors of
the mRNA distribution depend on the rate constants that control the
time
 scales of various processes. Therefore, we begin
 by briefly summarizing the different time scales that arise in the model.
  The model has four rate constants, the
forward and backward rates for the gene to switch between the active
and inactive states and the transcription and degradation rates that
govern mRNA numbers, leading to three independent, dimensionless
ratios.
 The equation obeyed by the probability for the DNA to
be in the excited state, denoted by $Q_1(t)$ can be obtained
directly from Equation (1):  $ d{Q_1}/dt\,=\, c_f\,-\,(c_f+c_b)\,Q_1
\,.$ This shows that the effective DNA relaxation rate to the steady
state is governed by $c_f+c_b$.  The mean mRNA number obeys the
equation  $d\langle m(t)\rangle/dt\,=\,-k_d\,\langle m
\rangle\,+\,k_b\,Q_1(t)$.  It is thus clear  that the temporal behavior
of the mean mRNA number is determined by the rates $k_d$ and
$c_f+c_b$, the latter entering since it determines the dynamics of
the transcriptionally active state. As long as $k_d\,<\,c_f+c_b$,
the mRNA decay rate sets the time scale over which relaxation to the
steady state occurs. We find that the results of our exact solution
can be interpreted in the most natural and transparent way when we
measure time in units of $k_d^{-1}$, i.e. in terms of the mRNA
lifetime. Thus we will use the three dimensionless ratios $k_b/k_d$,
$c_f/k_d$, and $c_b/k_d$ to organize our results.
The mean number of mRNA in the steady state is given by
the product of
 $c_f/(c_f+c_b)$, the fraction of the time the gene is in the activated
state and $k_b/k_d$, the mean value of mRNA if the gene is always
`on'. The ratio  $k_b/k_d$ clearly sets the scale for the number of
mRNA and increasing it extends the range over which $P(m)$ is
appreciable without a significant change of shape. The remaining
ratios $c_f/k_d$ and $c_b/k_d$
 determine the shape of the distribution.

\subsection{Superposition of Poisson distributions}~~
We begin by providing an intuitively appealing way to view our exact
result for $P(m,t)$.  The key conclusion is that $P(m,t)$
 can be pictured as a
superposition of Poisson distributions with different mean values.
If the gene is always ``on'' the mRNA distribution in steady state
is Poisson with the Poisson parameter, $\lambda$, given by the mean
$k_b/k_d$, the ratio of transcription and degradation rates. Since
the gene flips between the ``on" and ``off" states with the rates
determined by $c_f$ and $c_b$, the mRNA distribution is determined
by a stochastic transcription rate $k_b\zeta(t)$ where $\zeta(t)$ is
a dichotomous noise that assumes values $0$ or $1$ corresponding to
the gene being in the inactive or active state respectively. The
dynamics of the random variable $\zeta$ is determined by the
stochastic chemical reaction described by Equations (1). Thus the
distribution of the mRNA number is described by a Poisson process in
which the parameter $\lambda$ itself is stochastic, a process called
a doubly stochastic Poisson process~\cite{cox}.

Consider observing a particular cell at a time $T$. The number of
mRNA at time $T$ is distributed according to a Poisson distribution
with parameter $\lambda(T)$ that depends on the time history of
$\zeta(t)$ from $0$ to $T$ describing the sequence of flips between
the on and off states in that cell. This time history corresponds to
a series of pulses of unit height with both the widths of the pulses
and the intervals between pulses independently and exponentially
distributed with parameters $c_f$ and $c_b$ respectively. Different
cellular behaviors correspond to different realizations of the
random sequence of pulses. Thus cell to cell variability in mRNA
copy number is given by  a superposition of Poisson distributions
with parameter $\lambda$ that is itself a random variable: \beq P(m,
t) \,\,=\,\int d\lambda\,\rho(\lambda,
t)\,e^{-\lambda}\frac{\lambda^m}{m!} \eeq where $\rho(\lambda, t)$
is the probability density of the random variable $\lambda$.
 The fraction
of cells with $m$ copies of mRNA at time $t$ is determined by
$\rho(\lambda, t)$. Such superpositions have been considered in the
context of stochastic processes, for example, in \cite{gardiner}.
This representation provides an attractive conceptual framework for
understanding the transcriptional pulsing problem. We shall
elaborate elsewhere on how the different forms of $\rho(\lambda, t)$
in the different regions of parameter space allow us to interpret
the corresponding  behaviors of $P(m,t)$.

\subsection{Steady state distributions}
\begin{figure*}

\centering

\rotatebox{-90}{\resizebox{!}{10cm}{\includegraphics{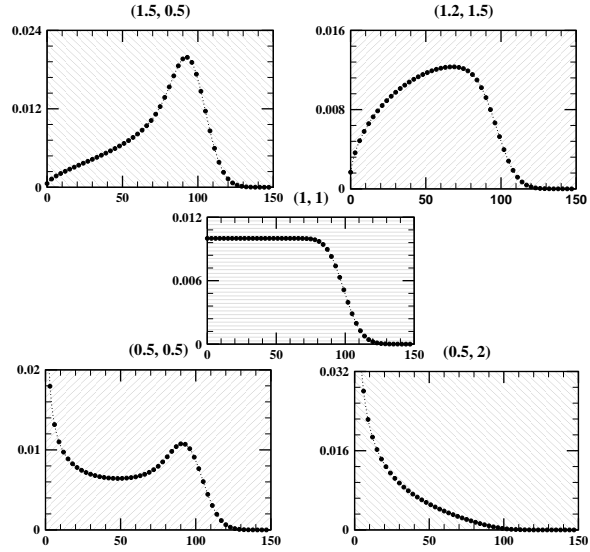}}}
\caption{ Steady state mRNA distributions $P(m)$ vs. $m$, labeled by
$(c_f, c_b)$ in units of $k_d$. The mRNA transcription rate $k_b$ is
$100 k_d$ for all the distributions. The figure shows prototype
distributions for the five major
 regimes $c_f, c_b \lessgtr k_d$ in parameter space. The distribution
for $c_f, c_b = k_d$ is flat. Bimodals are obtained in the lower
left panel for $c_f,c_b < k_d$ while a  power law occurs in the
lower right panel. } \label{map}
\end{figure*}

~~
We now describe the variety of steady state distributions that occur
in different regions of parameter space.
  In view of the
discussion of time scales in the previous section,
 it is natural to classify the distributions by plotting $c_f/k_d$
and $c_b/k_d$ along the $x$- and $y$-axes
  for fixed
  $k_b/k_d$. The results for fixed value $k_b/k_d\,=\,100$ are
displayed in Figure~\ref{map} and provide a bird's eye view of the
strikingly different mRNA distributions that arise in different
regions of parameter space. We recall that the experiments are
performed on a variety of organisms both prokaryotic and eukaryotic.
While rate constants are not known, given the different time scales
involved in the experiments, we have chosen to investigate a range
of values of $c_f/k_d$ and $c_b/k_d$ that encompass different
biologically significant cases: for example, our choices include the
vastly different rate constant values in the experiments of Raser
and O'Shea~\cite{raser} and Raj et al.~\cite{raj}.

We start with the interesting case displayed in the bottom left
figure in Figure~\ref{map}, when the mRNA half-life is
 shorter than the the time scales over which the gene
turns on or off, i.e.,  $c_f, c_b < k_d$.
 In the steady state at any
given time, the gene is off in some cells. Since the mean duration
of the pulse $1/c_b\,>\,1/k_d$ the transcripts produced in the
previous occurrence of the on state would probably have decayed and
so the number of transcripts will usually be small in these cells.
This causes a peak in the mRNA distribution near $m=0$. In those
cells in which the gene is on at the time of observation the number
of transcripts can display a broad range of values depending on how
long the gene was active as compared to the mRNA lifetime. Thus, we
expect to observe a bimodal distribution, as was qualitatively
argued in~\cite{hume}. The result is shown in the lower left
quadrant of Figure~\ref{map}. One finds a peak at $m = 0$ (with a
weight that can be computed analytically) and another peak at large
($\sim k_b/k_d$) $m$ values. If the values of $c_f$ and $c_b$ are
such that the peaks are well-separated, much of the intermediate
region displays a power-law behavior. This reflects the broad range
of times for which the gene has been active in different cells at
the time of observation. It is useful to remark that bimodal
distributions have been obtained in models with feedback~
\cite{ferrell}. In contrast, in the transcriptional pulsing model
bimodality is obtained \emph{without} the presence of a feedback
loop.

Now imagine that we keep $c_f$ fixed and vary $c_b$ so that it is
larger than the decay rate. This leads to a scale-invariant,
power-law behavior over a significant range of mRNA values. This
simple power-law decay obtained in the case $c_f< k_d$ and $c_b >
k_d$ is illustrated in the lower right quadrant of Figure~\ref{map}.
This case has been treated analytically
 in~\cite{bose, raj}. The mRNA distribution  can be fitted
by a Gamma distribution, which for appropriate values of the rate
constants, shows power-law behavior over a substantial range.

When both the activation and inactivation rates are rapid, i.e.,
$c_f, c_b \gg k_d$, eliminating the fast reactions naively yields a
simple birth-death process for the mRNA with an effective
transcription rate $k_b\times c_f /(c_f + c_b)$. This would lead one
to expect a Poisson distribution for the mRNA number. However, in
this `quadrant', i.e. for $c_f, c_b
> k_d$, the observed distribution has a broad single-humped shape as
displayed in the upper right quadrant of Figure~\ref{map}, much
broader than a Poisson distribution. This broadening occurs because
the parameter $\lambda$ itself is stochastic.  When $c_f
>k_d> c_b$ the gene is on most of the time. Not surprisingly, the
distribution is Poisson to a very good approximation as seen in the
upper left quadrant of Figure~\ref{map}. In the intermediate region
when $c_f, c_b \sim k_d$ the distribution interpolates between these
different possibilities and is rectangular when they are equal (see
Figure~\ref{map}, center).

\subsection{Time evolution of probability distributions}
\begin{figure}
 \centering
{\resizebox{!}{6.8cm}{\includegraphics{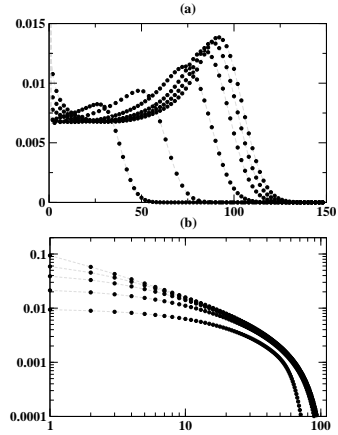}}} \caption
 {Time evolution of $P(m, t)$ towards steady state as a function of $m$
at different time points
(a) in the
bimodal regime for
$c_f=0.75 k_d$, $c_b=0.5 k_d$ and $k_b=100 k_d$ at times $t = 0.5$,
$1$, $2$, $3$, $4$ and $\infty$ in units of $k_d^{-1}$. The steady
state with bimodality is reached around $4k_d^{-1}$. (b) Log-Log
plot of $P(m, t)$ for $c_f=0.25 k_d$,
$c_b=2.5 k_d$ and $k_b=100 k_d$ at times  $t = 1$, $2$, $3$, $4$ and
$\infty$ in units of $k_d^{-1}$. The figure shows the time evolution
of the power law region of $P(m, t)$. For the different curves time
increases from bottom to top with the slope increasing to its
steady-state value.
  }
\label{bimodaltime}
\end{figure}

~~
Because of the range of possible time scales, it can happen that that the time
when measurements are made, the biological system has not attained steady state.
For this reason, we now present results for how the distributions evolve to a steady
state from an initial state with no mRNA and the gene in its inactive state.
 Using the time-dependent result for the distribution (see Material and Methods, equation
(9)), we evaluate the evolution using
\emph{Mathematica} and plot the complete probability distribution as
a function of time. Consider the case $c_f, c_b < k_d$,  (bottom
left in Figure~\ref{map}),
 where the mRNA distribution displays
bimodality. Here the mRNA decay rate sets the scale for approach to
the steady state. Figure~\ref{bimodaltime}(a) shows the evolution of
the bimodal distribution as a function of time. For the given
initial condition the second peak away from zero develops after a
period of roughly twice the mRNA half-life. Steady state behavior
sets in at about 4 to 5 times $k_d^{-1}$.    It is clearly possible
that, depending on the relative values of the cell cycle time and
the mRNA half-life, steady state and therefore, full bimodality may
not be observable.

Consider now the time evolution of the distribution that evolves
into the``pure'' power law behavior featured in the bottom right of
Figure~\ref{map}. In Figure~\ref{bimodaltime}(b) we plot $P(m,t)$ vs
$m$ on a double logarithmic plot.  We have chosen $c_f=0.25k_d$ and
$c_b=2.5k_d$ to illustrate this case. Larger values of the
transcription rate will lead to a larger range over which the power
law behavior obtains. It is clear that the exponent of the power law
increases in magnitude with time and saturates at the steady state
value for $t$ greater than about $4k_d^{-1}$. Thus the shape the
distribution depends crucially on the time (measured in units of the
decay time) when experimental measurements are made.


\subsection{Mean and variance versus full distribution}~~
From the examples given in Figure 1 it is clear that the complete probability distribution
of mRNA number is required to characterize the
behavior of the transcriptional pulsing model. Nevertheless, for
completeness, we make some remarks concerning attempts to represent
a mRNA distribution by its mean and variance only.

 We recall the expressions for the mean and variance in
mRNA number in the transcriptional pulsing model reported in
~\cite{raser}. The mean is given by $\mu\,=\,k_bc_f/(k_d(c_f+c_b))$
and the variance by  \beq \sigma^2 = \frac{k_b}{k_d}\frac{ c_f}{c_f
+ c_b}\left( 1+ \frac{k_b c_b}{(c_f + c_b)(k_d
 + c_f + c_b)}\right)
\eeq The first term in (5) is equal to the mean while the second
term arises from the stochasticity in the pulsing process. It can be
shown~\cite{cox} that \beq \sigma^2 = \langle \lambda \rangle +
\sigma_{\lambda}^2 \eeq Thus, in a doubly stochastic birth-death
process the variance in mRNA number has an additional contribution
due to the stochasticity of gene activation and inactivation.

\begin{figure}
 \centering
\resizebox{!}{4.8cm}{\includegraphics{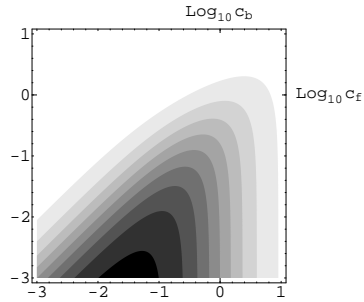}} \caption
  {Contour plot of the noise strength(Fano factor) $\eta$ as
$c_b$ and $c_f$ (in units of $k_d$) are varied, for $k_b =1000 k_d
$, in the steady state;
  $c_b$ and $c_f$ are varied on a $\log_{10}$ scale over $5$
  decades.
  $9$ contours for different values of $\eta$ are placed at intervals of $100$, from
  $1$ to $1001$ with
  $\eta$ increasing from light to dark values.
}
 \label{eta_contours}
\end{figure}

There are two popular measures of noise in terms of the first two
moments of a probability distribution: the coefficient of noise, $\xi$, defined as the ratio
of
the standard deviation $\sigma$ to the mean $\mu$, and the noise
strength or Fano factor,  $\eta$, defined as  the ratio of the
variance $\sigma^2$ to the mean . The latter has the value of unity
for a Poisson distribution and is therefore convenient for
describing deviations from Poisson behavior. In
Figure~\ref{eta_contours} we display constant $\eta$ contours as a
function of $c_f/k_d$ and $c_b/k_d$ for a fixed value of $k_b/k_d$
on a logarithmic scale to encompass a broad range of parameter
variation. When $c_f < k_d$ and  $c_b
> k_d$, the steady state distribution $P(m)$ is monotonically decreasing and
has a power law region. In this region, to a first approximation
$\eta$ is independent of $c_f$ (and $ \approx 1 + k_b/(k_d+c_b)$)
and the contours are roughly parallel to the $c_f$ axis. This
emphasizes the possibility that $\sigma^2/\mu$ is a constant for
systems with power-law behavior in which $c_f$ varies over a broad
range of values. Since as we show later, the protein distribution
can reflect the behavior of the corresponding mRNA distribution, the
protein distribution can show a similar constancy of the Fano
factor.  Such a behavior has been observed experimentally in
~\cite{barkai} where a pulsing model was discussed. In the region
where  $c_f, c_b < k_d$, then $\eta \approx 1 + (k_b/k_d)/(1+
c_f/c_b)$ and thus depends only on $c_f/c_b$. This is consistent
with the contours in this region being straight lines with slope
$1$. For the region with $c_f, c_b>k_d$, there is rapid switching
between on and off states and the Fano factor depends weakly on the
rates $c_f$ and $c_b$.

\begin{figure*}
\centering
\rotatebox{-90}{\resizebox{!}{8.0cm}{\includegraphics{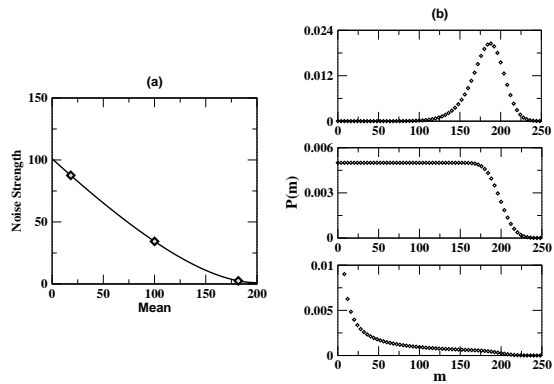}}}
\caption{ Smoothly varying noise strengths as activation rate is
varied can correspond to different probability distributions. (a)
Variation of noise strength (Fano factor) with activation rate for
$k_b = 200 k_d$ and $c_b = k_d$ (b) The steady state distributions
$P(m)$ corresponding to the (diamond-shaped) points marked in (a).
The points in (a) going from right to left and the corresponding
figures from top to bottom in (b) are for $c_f/k_d = 10, 1$, and
$0.1$ respectively. Figure (b) illustrates how different points on
the same curve (a) can be associated with dramatically different
mRNA distributions.} \label{raser}
\end{figure*}

There is danger in characterizing distributions solely by their mean
and variance. The variety of possible mRNA distributions across
cells shown in Figure~\ref{map}, demonstrates this. One of the
interesting results in Reference \cite{raser} showed the decrease in
the noise strength (Fano Factor) with increase in the mean for genes
with different activation rates. Here we
show that a wide variety of distributions underlies this
correlation between the noise strength and the mean. The increase in
the mean can be obtained in the model through an increase in the
activating rate, namely the rate $c_f$, and experimentally through
mutations of an appropriate promoter~\cite{raser}. Even though a
smooth curve is obtained for the decrease of noise strength with the
mean, we illustrate how the full mRNA distribution can differ for
different points along the curve. For specificity, we choose
parameter values $k_b\,=\,200 k_d$ and $c_b\,=\,k_d$, and vary the
forward rate $c_f$ for gene activation which changes the mean value.
The result is shown in Figure~\ref{raser}(a) and is similar to that
obtained experimentally.  Now we examine the full probability
distribution at three values of $c_f$, namely
$c_f\,=\,0.1k_d,\,k_d$, and $10 k_d$, which correspond to mean
values of $18$, $100$, and $181$ respectively. The mRNA
distributions are shown in Figure~\ref{raser}(b): the
distribution ranges from power-law decay of $P(m)$ for $c_f=0.1k_d$
to a broadened Poisson distribution for $c_f=10k_d$. Furthermore, as
we pointed out earlier, the value of mRNA degradation rate plays an
important role in determining the type of mRNA distribution, a role
not apparent in the regimes discussed in~\cite{raser}.

Since we have an analytic expression for the complete distribution
we can use an information-theoretic characterization of the mRNA
probability distribution, the Shannon entropy. We evaluate the
Shannon entropy for different values of the rate constants. While
the flat distribution clearly corresponds to high entropy, the
power-law distribution also yields large values of the entropy
indicating that greater information content than the other
distributions. The results are presented in Supplementary Section B.

\subsection{Protein distributions}~~
Given the variety of mRNA distributions that
can result from genes undergoing transcriptional pulsing, it is
important to understand how this affects the probability
distributions for the corresponding protein. While a careful answer
to this question would require detailed modeling of mRNA
translocation and translation, we address this issue in a model in
which translation is treated as a Poisson process~\cite{thattai}.
Thus, we use \beqa
M&\stackrel{p_b}{\longrightarrow}& M \,+\,P \\
P &\stackrel{p_d}{\longrightarrow}&\emptyset \,.\eeqa The effective
protein degradation rate would include contributions from cell
division, dimerization and other gene specific processes involving
the loss of proteins.  From our results we identify two regimes, one
in which the protein distribution is similar to the mRNA
distribution and another in which they are different.

\begin{figure}
\centering {\resizebox{!}{6.8cm}{\includegraphics{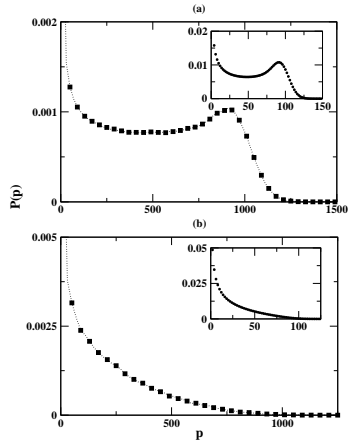}}}
\caption{Examples of steady state distributions of proteins
reflecting those of mRNA.  Protein distributions (a) for
$c_f=0.5 k_d$, $c_b=0.5 k_d$ and $k_b=100 k_d$, $p_b = 20 k_d$ and
$p_d = 2 k_d$ (b) for $c_f=0.5 k_d$, $c_b=2 k_d$ and $k_b=100 k_d$,
$p_b = 20 k_d$ and $p_d = 2 k_d$. The insets show the corresponding
mRNA distributions. When the protein degrades faster than the mRNA,
its distribution qualitatively mirrors the corresponding mRNA
distribution. } \label{MP_same}
\end{figure}

The protein distribution qualitatively mirrors the mRNA distribution
if the protein dynamics is faster than the mRNA dynamics:
translation and protein degradation occur at a rate higher than the
 mRNA degradation rate $k_d$ (with $c_f + c_b$ of the order of
$k_d$). In this case, bimodals give rise to bimodals, power-laws
give rise to power-laws and so on. The results are displayed in
Figure~\ref{MP_same}, where the protein distributions for two of the
cases illustrated in Figure~\ref{map} are shown along with the
corresponding mRNA distributions. We show results for the protein
degradation rate $p_d$ set to twice $k_d$, the mRNA degradation
rate. Figure~\ref{MP_same}(a) has rate constants $c_f, c_b$ such
that the mRNA distribution exhibits bimodality, as seen in the
inset. The protein distribution is also clearly bimodal for this
case. Similarly, in Figure~\ref{MP_same}(b), a long tailed mRNA
distribution with a power law region gives rise to a similar protein
distribution when the other rate constants are appropriately chosen.

\begin{figure}
\centering {\resizebox{!}{6.8cm}{\includegraphics{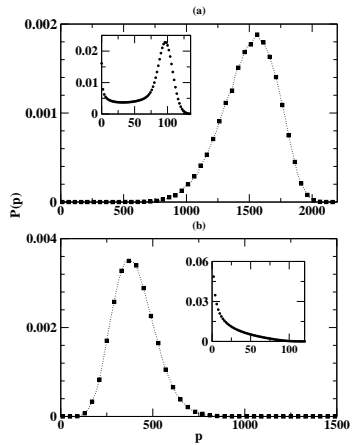}}}
\caption{Examples of steady state distribution of proteins
differing from those of mRNA.  Protein distributions for (a) for
$c_f=0.6 k_d$, $c_b=0.2 k_d$ and $k_b=100 k_d$, $p_b = 1 k_d$ and
$p_d = 0.05 k_d$ (b) for $c_f=0.5 k_d$, $c_b=2 k_d$ and $k_b=100
k_d$, $p_b = 1 k_d$. The protein lifetime is $20$ times longer than
the mRNA lifetime.  The insets show the corresponding mRNA
distributions. When the protein degrades slowly compared to the
mRNA, its distribution can be qualitatively different from the
corresponding mRNA distribution.} \label{MP_diff}
\end{figure}

The other important case is when the proteins are relatively stable
compared to the mRNA. This case is more complex; however, there are
ranges of rate constants for which the protein and mRNA
distributions are vastly different. We illustrate this with two
examples.  If the rate constants are chosen such that the mRNA
distribution is bimodal (cf. Figure 1), then $P(p)$ is qualitatively
different from $P(m)$ if $p_d < c_f + c_b$. The protein
distributions for this case may be either monotonically decreasing
or bell-shaped, depending on whether $p_d$ is less than one or both
of $c_f, c_b$. This contrast between the protein and the mRNA
distributions is illustrated in Figure~\ref{MP_diff}(a). A second
case is displayed in Figure 6(b) for values of the rate constants
that lead to a power-law distribution for the mRNA. Here a
bell-shaped distribution is obtained for $P(p)$ when $p_d << c_f <
k_d<c_b$. These examples demonstrate that one has to be careful in
inferring the shape of one of the mRNA or protein distributions from
the other.

It is worth noting that even if mRNA numbers are small, even as low as
$\mathcal{O} (10)$, the above conclusions continue to hold. Thus, a
wide variety of protein distributions that may or may not reflect
the underlying mRNA distribution could be realized in real
biological systems when the gene undergoes transcriptional pulsing.

It has been argued recently~\cite{sunney}  that if the effective
protein degradation is very slow compared to that of the mRNA, the
protein distribution can be approximated by a gamma distribution.
While gamma distributions do provide a good fit for some regions of
parameter space, none of the distributions obtained in the bimodal
quadrant can be reasonably approximated by a gamma distribution.



\section{Discussion}
In this paper, we have presented results for the time-dependent and
steady-state probability distributions for mRNA in a transcriptional
pulsing model. A variety of mRNA distributions occur in different
regimes of rate constants. Our aim is to provide a guide for the
interpretation of data on cell-to-cell variability that could arise
from transcriptional pulsing.   Transcriptional pulsing, entailed by
the dynamics of chromatin remodeling, reinitiation and similar
processes, appears as a straightforward mechanism leading to
bimodality and also to mRNA distributions with long tails.
Long-tailed distributions of mRNA have been seen in a variety of
systems: the experiments of Raser and O'Shea~\cite{raser} show
evidence for long tails which they attribute to transcriptional
pulsing. In experiments on the gene ActB in cells from mouse
pancreatic islets the distribution of mRNA number, $m$, was found to
be consistent with a log-normal distribution that would correspond
to $P(m)\,\sim\,m^{-1}$ over some range of $m$~\cite{bengtsson}.
More recently, a similar distribution of the Interferon-$\beta$ gene
transcripts has been found in human dendritic cells~\cite{hu}. For
the latter two experiments our results should help clarify whether
the origin of the mRNA behavior lies in transcriptional pulsing.
However, cell behavior is controlled not by mRNA but by the proteins
they encode. Therefore, it is crucial to determine whether the
protein distribution follows the corresponding mRNA distribution. We
have determined when the two distributions are similar but also
identified situations when the protein distributions are
strikingly different from those of the mRNA. In general, and for
these situations in particular, our results on the range of
cell-to-cell variability of mRNA and protein responses due to
transcriptional pulsing should provide significant help in
interpreting experiments.

\section{Materials and Methods}
The results presented and discussed here for mRNA are based
on the exact form of the distribution function $P(m,t)$ of the mRNA
number, $m$ at time $t$. We have solved the Master
Equation~\cite{gardiner} for reactions (1)-(3) that describes the time evolution of the
distribution to obtain these results. We found it convenient to work
with the generating function defined by
$G(z,t)\,=\,\sum_{m=0}^\infty\,z^m\,P(m,t)$. If we can evaluate
$G(z,t)$ exactly, then the probability of having $m$ mRNA
transcripts at time $t$ can be obtained by extracting the
coefficient of the $z^m$ term.  We have computed the generating
function exactly (See Supplementary Section A for the details) for
the initial condition with zero mRNA, i.e., $P(m,0)=\delta_{m,0}$,
and find
 \beq
 \begin{split}
&G(z,t) = F_s(t) \Phi(c_f,c_f+c_b;-k_b(1-z))  \\
&+ F_{ns}(t)(1-z)\Phi(1-c_b,2-c_f-c_b;-k_b(1-z))
\end{split}
\eeq where $\Phi$ is the (Kummer) confluent hypergeometric
function~\cite{abramowitz, slater} and the coefficients $F_s(t)$ and
$F_{ns}(t)$ can be calculated explicitly in terms of confluent
hypergeometric functions. The results are displayed in Supplementary
Section A. At large times $F_s\,=\,1$ and $F_{ns}\,=\,0$. Thus in
the steady state the generating function is given by( see also
~\cite{raj}) \beq G_s(z)\,=\,\Phi(c_f,c_f+c_b;-k_b(1-z))\,, \eeq

We can use the exact solution in Equation (9) to extract the
\emph{time-dependent} behavior of $P(m,t)$ or Equation (10) for the
steady state for different ranges of values of the rate constants.
The results presented were obtained by extracting the coefficients
of $z^m$ in the expansion of
 the generating function using \emph{Mathematica}\cite{wolf}.
 While standard numerical simulations based on the Gillespie
algorithm~\cite{gillespie} can be employed to study both the steady
state and the time evolution, the
 exact solution allows us to extract the results much
more efficiently and explore the behavior of the system
systematically in the space of rate constants without statistical
errors, especially, when there are long tails present. The results
for the protein distributions were obtained from numerical
simulations using the Gillespie algorithm.

\begin{acknowledgments}
This work was supported by the National Institute of Allergy and Infectious
Diseases through contract HHSN266200500021C.  We thank 
Stuart Sealfon, James Wetmur, and Jianzhong Hu for
discussions that stimulated this work and Stuart Sealfon for a
careful reading of the manuscript and comments. S. I. B. thanks Rudro R. Biswas for productive discussions.\\
\end{acknowledgments}









\end{article}









\end{document}


\begin{centering}
\textbf{Supplementary Section A\\}
\textbf{Time-dependent solution to the Master Equation for
transcriptional bursts}\\
\end{centering}

For the set of reactions described by Equations $1$-$3$ in the text
we define $P_0(m,t)$ and $P_1(m,t)$ to be the probability that at
time $t$ the cell has $m$ mRNA molecules and the gene is in the
inactive and active states respectively. It is straightforward to
write down the Master Equation for the two probabilities: \beqa
\frac{dP_0(m,t)}{d t}&=&
-c_fP_0(m,t)\,+\,c_bP_1(m,t)\,+\,k_d\,[(m+1)P_0(m+1,t)-mP_0(m,t)\,]\\
\frac{d P_1(m,t)}{dt}&=&
c_fP_0(m,t)\,-\,c_bP_1(m,t)\,+\,k_d\,[(m+1)P_1(m+1,t)-mP_1(m,t)\,] \nonumber \\
&& +k_b\,[\,P_1(m-1,t)\,-\,P_1(m,t)\,]
 \eeqa

 We define the generating functions
 $$G_\alpha(z,t)\,\equiv\,\sum_{m=0}^\infty\,z^m\, P_\alpha(m,t)\,$$ for $\alpha\,=\,0$ and $1$.
 The mRNA distribution (independent of the state of the gene) is determined by the sum $G\,\equiv\,G_0+G_1$.
 It is easy to deduce the equations obeyed
 by the generating functions from the Master Equations(with time re-scaled by $k_d$):
\beqa
\partial_tG_0(z,t)&=&-c_fG_0(z,t)\,+\,c_bG_1(z,t)\,+\,(1-z)\,\partial_zG_0(z,t)
\\
\partial_tG_1(z,t)&=&c_fG_0(z,t)\,-\,c_bG_1(z,t)\,+\,(1-z)\,\partial_zG_1(z,t)\,-\,
k_b(1-z)G_1(z,t).\eeqa All the rate constants are measured in units
of $k_d$. \\

We simplify the equations using an analog of the Galilean
transformation by making the change of variables
$v\,\equiv\,k_b(1-z)$ and
$w\,\equiv\,ve^{-t}\,=\,k_b(1-z)\,e^{-t}$. In terms of the
transformed variables, we have  \beqa \label{G0} v\partial_v
G_0&=&-c_fG_0\,+\,c_bG_1  \\
\label{G1} v\partial_vG_1&=&c_fG_0\,-\,c_bG_1-vG_1 \,. \eeqa Adding
the two equations we have the useful relation \beq \label{useful}
\partial_v(G_0+G_1)\,=\,-G_1\,\,.\eeq
Note that $G_0(z,t)$ and $G_1(z,t)$ (and hence, their sum) are
functions of $v$ only and independent of $w=k_b(1-z)e^{-t}$; the
dependence on $w$  is determined by the boundary conditions.

It is convenient to derive a second-order differential equation for
$G$. Therefore we  differentiate the equations for $G_0$ and $G_1$
and obtian (\ref{useful}) \beqa
v\partial_v^2G_0\,+\,(1+c_f+c_b+v)\partial_vG_0\,+c_fG_0&=&0 \\
v\partial_v^2G_1\,+\,(1+c_f+c_b+v)\partial_vG_1\,+\,(1+c_f)G_1&=&0\,.
\eeqa

We add the two equations and use Equation (\ref{useful}) to obtain
$$v\partial_v^2G\,+\,(c_f+c_b+v)\partial_vG\,+\,c_fG\,=\,0\,.$$
The substitution $G(v)=e^{-v}F(v)$ shows that $F(v)$ satisfies the
confluent hypergeometric equation in the canonical form. The
solution is given by \beq
F\,=\,A(w)\,\Phi(c_b,c_f+c_b;v)\,+\,B_0(w)\,v^{1-c_f-c_b}\,\Phi(1-c_f,2-c_f-c_b;v)\,\,.\eeq
Upon using the Kummer transformation,
$e^{-v}\Phi(\alpha,\gamma;v)\,=\,\Phi(\gamma-\alpha,\gamma;-v)$, we
obtain \beq \label{form1}
G\,=\,A(w)\Phi(c_f,c_f+c_b;-v)\,+\,B_0(w)v^{1-c_f-c_b}\Phi(1-c_b,2-c_f-c_b;-v)\,.\eeq
In order to obtain a well-defined power series in $v=k_b(1-z)$ for
the generating function we must impose
$$B_0(w)\,=\,w^{c_f+c_b}B(w)\,=\,v^{c_f+c_b}\,e^{-(c_f+c_b)t}\,B(w)\,.$$
This yields the form \beq \label{form2}
G\,=\,A(w)\,\Phi(c_f,c_f+c_b;-v)\,+\,B(w)\,e^{-(c_f+c_b)t}\,v\Phi(1-c_b,2-c_f-c_b;-v)
\,.\eeq

 We impose the boundary conditions at $t=0$ which corresponds to $w=v$. The initial condition
  $P(m,t=0)\,=\,\delta_{m,0}$
 leads to \beq G(w=v,v)\,=\,1 \,.\eeq
We assume that the gene is initially in the inactive state and thus
$G_1(z,t=0)=0$. The additional condition that arises from Equation
(\ref{useful})implies \beq
\partial_v G(w,v)\vert_{w=v}\,=\,0\,.\eeq
Imposing these conditions we determine the unknown functions $A$ and
$B_0$. This involves judicious use of the
 Wronskian identity \beq
\label{newwron}
\Phi(\alpha-\gamma+1,1-\gamma;z)\Phi(\alpha,\gamma;z)
-\,\frac{\alpha}{\gamma(1-\gamma)}
z\Phi(\alpha-\gamma+1,1-\gamma;z)\Phi(\alpha+1,\gamma+1,z)\,=\,
e^{z}\,\eeq that follows from from results in Ref.~\cite{lucy} and
other identities to found there. The final result is
 \beq
G(z,t)\,=\,F_s(t)\,\Phi(c_f,c_f+c_b;-k_b(1-z))\,
 +\,F_{ns}(t)\,(1-z)
\,\Phi(1-c_b,2-c_f-c_b;-k_b(1-z))\,.\eeq where \beqa F_s(t)&=&
\Phi(-c_f,1-c_f-c_b;k_be^{-t}(1-z))~~~~\mbox{and}\\
F_{ns}(t)&=&-\frac{c_fk_b(1-z)}{(c_f+c_b)(1-c_f-c_b)}
\,e^{-(c_f+c_b)t} \Phi(c_b,1+c_f+c_b;k_be^{-t}(1-z))\,.\eeqa

\begin{centering}
\textbf{Supplementary Section B\\}
\textbf{Characterization of noise in terms of Shannon Entropy} \\
\end{centering}

In information theory the Shannon entropy serves as a measure of the
average information content or the uncertainty of a random variable.
We now present results for this  measure of the noise in the mRNA
distribution. Given the exact solution we can directly evaluate the
Shannon entropy associated with the steady state distribution $P(m)$
defined by \beq \label{shannon} \mathcal{S} = -\sum_{i} P(m)\log_{2}
P(m) \,.\eeq

\begin{figure}
\label{Shannon1} \centering
\rotatebox{-90}{\resizebox{!}{12cm}{\includegraphics{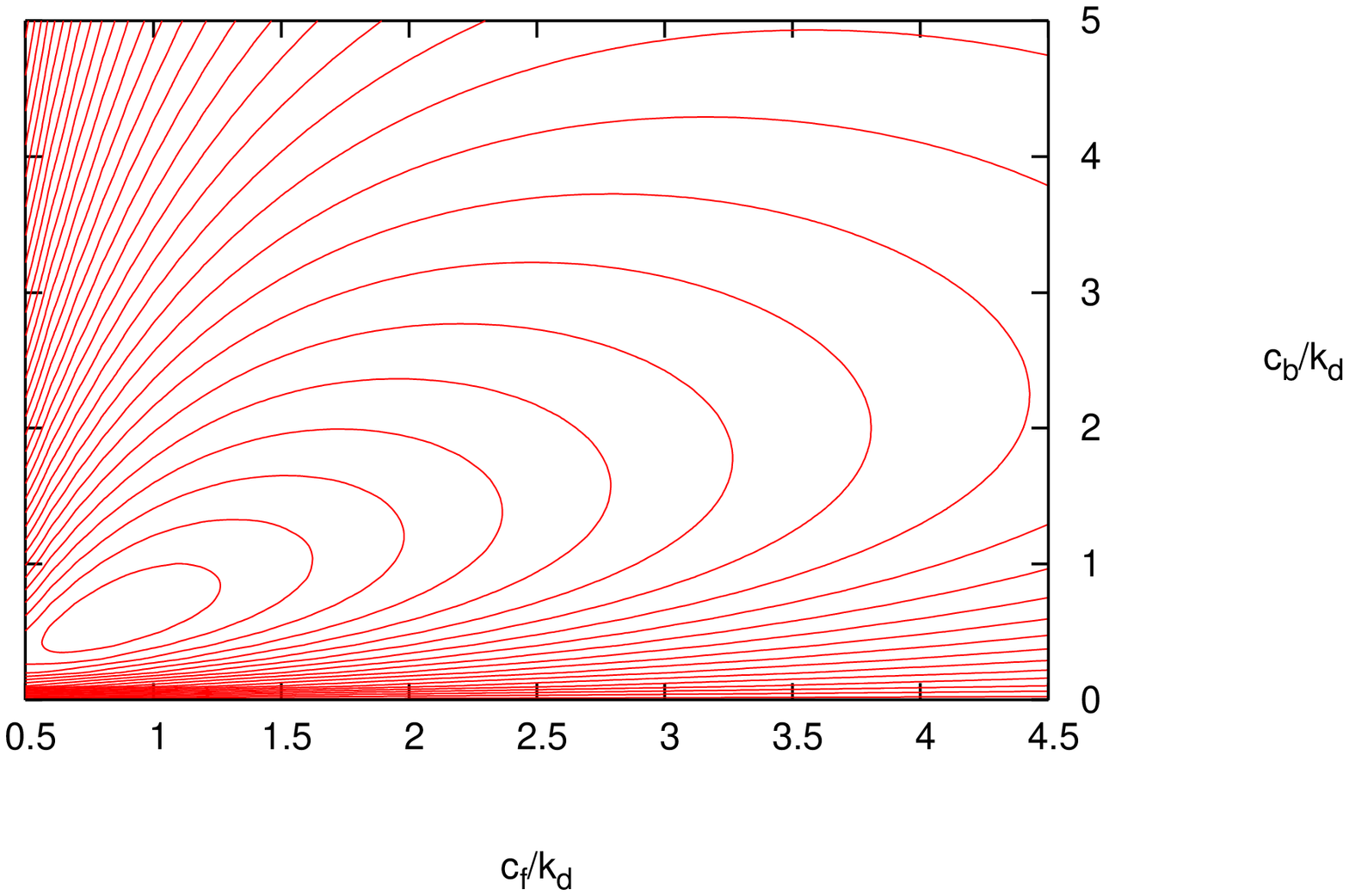}}}
\caption{ Contour Plot of the Shannon entropy  defined in
Equation~\ref{shannon} of $P(m)$ for $k_b = 100 k_d$; separation between
contours is $0.05$ and the Shannon entropy decreases outward from
the central contour. }
\end{figure}

 In Figure 1~\ref{Shannon1} we display contours of constant entropy as a
 function of the forward and backward rates $c_f$ and $c_b$.  Not surprisingly,
 the values
$c_f/k_d, c_b/k_d \simeq 1$ yield the largest entropy since this
choice leads to a uniform distribution in $P(m)$. The power law
distribution also provides a range of values for the rates in which
larger values of entropy can be obtained. Since the Shannon entropy
is a measure of the amount of information required to describe the
random variable on average it can be helpful in the interpretation
of data. For example, consider dendritic cells involved in providing
innate immunity to an organism against pathogens. Assume that the
mRNA or the protein produced by the cell to overcome a viral
antagonist has a broad distribution. It is plausible that the
greater the amount of information required to describe the
distribution the lesser the chances of the pathogen being able to
overcome the organism's immune system by random mutations. This can
confer greater immunity against mutations in the virus that could
evade the defence mechanisms of the organism. It is known that the
interferon-$\beta$ mRNA distribution is broad in human dendritic
cells \cite{hu}.